\documentclass[]{emulateapj}
\slugcomment{accepted for publication in Ap.J. Letters}

\shorttitle{Planetary Companion to HD 37605}
\shortauthors{Cochran et al.}

\begin{document}

\title{The First HET Planet: A Companion to HD 37605 
\footnote{Based on observations obtained with the Hobby-Eberly Telescope, which
is a joint project of the University of Texas at Austin, the Pennsylvania State
University, Stanford University, Ludwig Maximilians Universit\"{a}t
M\"{u}nchen, and Georg August Universit\"{a}t G\"{o}ttingen.}}

\author{William D. Cochran, Michael Endl, Barbara McArthur}
\affil{McDonald Observatory, University of Texas at Austin, Austin, TX 78712}
\email{wdc@astro.as.utexas.edu, mike@astro.as.utexas.edu, mca@barney.as.utexas.edu}

\author{Diane B. Paulson}
\affil{Department of Astronomy, University of Michigan,  Ann Arbor, MI 48109}
\email{apodis@umich.edu}

\author{Verne V. Smith}
\affil{Department of Physics, University of Texas at El Paso, El Paso, TX 79968}
\email{verne@barium.physics.utep.edu}

\author{Phillip J. MacQueen, Robert G. Tull, John Good,
John Booth, Matthew Shetrone,
Brian Roman, Stephen Odewahn, Frank Deglman, Michelle Graver, Michael Soukup,
Martin L. Villarreal Jr.}
\affil{McDonald Observatory, University of Texas at Austin, Austin, TX 78712}
\email{pjm@wairau.as.utexas.edu,
rgt@astro.as.utexas.edu,
good@astro.as.utexas.edu,
booth@astro.as.utexas.edu,
shetrone@shamhat.as.utexas.edu,
bert@mcs.as.utexas.edu,
sco@astro.as.utexas.edu,
deglman@astro.as.utexas.edu,
michelle@astro.as.utexas.edu,
msoukup@astro.as.utexas.edu,
mlv@astro.as.utexas.edu}

\begin{abstract}
We report the first detection of a planetary-mass companion to a star
using the High Resolution Spectrograph (HRS) of the Hobby-Eberly Telescope
(HET).  The HET-HRS now gives routine radial velocity precision of
2-3\,m\,s$^{-1}$ for high SNR observations of quiescent stars.
The planetary-mass companion to the metal-rich K0V star HD\,37605
has an orbital period of 54.23~days, an orbital eccentricity of 0.737,
and a minimum mass of 2.84~Jupiter masses.  
The queue-scheduled operation of the Hobby-Eberly Telescope enabled us to
discovery of this relatively short-period planet with a total observation
time span of just two orbital periods.
The ability of queue-scheduled large-aperture telescopes to respond quickly
to interesting and important results demonstrates the power of this new
approach in searching for extra-solar planets as well as in other ares of
research requiring rapid response time critical observations.

\end{abstract}


\section{Introduction}
Traditional ground based radial velocity searches for extra-solar planetary
systems have used medium- and large-size telescopes assigned to the program
for particular nights of observing.
While this type of approach has been extremely
successful, resulting in about 120 detections of planetary-mass companions to
solar-type stars, the vagaries of the telescope scheduling process
introduces certain restrictions and constraints on the optimal acquisition
of time-critical observations.   In radial velocity surveys, it is often the
case that once a candidate planetary system has been identified and a
preliminary orbit determined, additional observations must be obtained at
particular orbital phases in order to place tighter constraints on particular
orbital elements.   Obtaining the necessary data at the optimal orbital
phase may be extremely difficult if observing runs are scheduled for a few
nights once per month, or even less frequently.  
Thus a telescope that is queue scheduled, such as the Hobby-Eberly Telescope,
(HET)  permits the observer to react quickly to the data and to obtain timely
follow-up observations of critical targets at the orbital phases where
they are needed.  We report use of the HET to detect a
planetary companion in a 54.23\,day period orbit to the star HD\,37605.
The ability of the HET observing queue to accommodate our observing needs
enabled us to discover the planet with only a total of 100 days (slightly less
than two orbital periods) elapsing between the first and final observations.
We discuss both the detection of the planetary companion itself and the
special attributes of the HET that enabled the final orbital confirmation to be
obtained in a very timely and expeditious manner.

\section{The Hobby-Eberly Telescope and its High Resolution Spectrograph}
The Hobby-Eberly Telescope consists of an array of 91 hexagonal
shape, spherical figure mirrors, each of 1\,meter diameter,
in a fixed truss tilted at $35^\circ$ zenith distance, resulting in a maximum
effective aperture of 9.2\,m diameter.
Declination ranges in the sky between $-11^\circ$ and $+72^\circ$ may be
selected by rotating the telescope truss in azimuth.
A star is observed by tracking an optical fiber bundle along the telescope
focal plane while the telescope structure remains fixed, and feeding the light
into the spectrograph.
The HET design was optimized for large-scale, low sky-density spectroscopic
surveys.  The telescope is {\it not\/} assigned to a given program for a given
night, but rather is queue scheduled, thus interleaving
observations from a large number of different programs employing different
instrument configurations.  The HET
data distribution policy is to allow the PI to gain access to the 
pervious nights data every morning allowing rapid feed back to the 
HET astronomers.
This is an ideal mode of operation for our high-precision radial velocity
program.

The High Resolution Spectrograph (HRS) \citep{Tu98} for the HET
was built by a team headed by Robert Tull and Phillip MacQueen.
This spectrograph was designed to be able to make stellar radial velocity
variation measurements with a precision of 3\,m\,s$^{-1}$ or better
on stars as faint as ${\rm V} = 10$.
The instrument is housed in an insulated enclosure in the basement
of the HET building, and thus is not moving with the telescope.
The image scrambling provided by the 34.3\,m long
optical fiber feed helps ensure that the pupil seen by the spectrograph
is always illuminated in a consistent manner.
Only a set number of instrument configurations are available,
and these are all defined by kinematic mounts
for the grating and cross-disperser positions.
The I$_2$ gas cell is contained in a collimated section
in front of the spectrograph slit, so refocusing of any optics is
unnecessary for I$_2$ cell use.
The detector is a mosaic of two $2048 \times 4100$ pixel E2V CCDs.
For our high-precision radial velocity observations, we use the HRS in
a mode giving resolving power of 60,000.   The echelle grating was used
on the center of the blaze, and the cross disperser was positioned so
that the break between the ``red'' and the ``blue'' CCD chips was
at around 5940{\AA}.  For most seeing conditions, a 2\,arcsec diameter
optical fiber was used.

\begin{figure}
\plotone{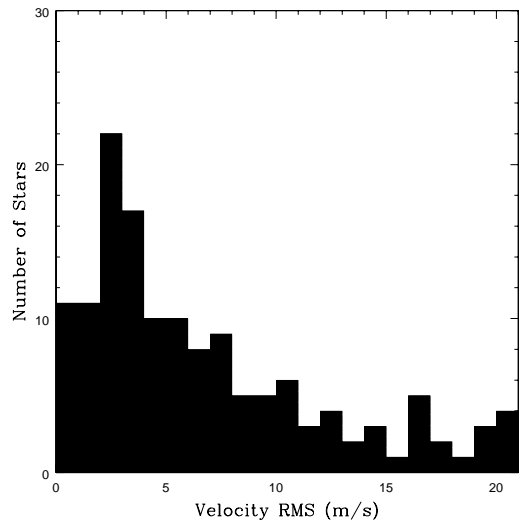}
\caption{Histogram of the velocity rms of 142 stars observed in
this program with the Hobby-Eberly Telescope High Resolution Spectrograph.}
\label{het_rms}
\end{figure}

So far, we have observed 173 F-M dwarfs four or more times.
The F through K stars are selected to be stars with low levels of stellar
activity based on X-ray emission, photometric
variability, and measured Ca~II H\&K emission.
We intentionally do not bias the selection on stellar metallicity, but
we do attempt to exclude stars with $v \sin i$ greater than about 
15\,km\,s$^{-1}$ and known short-period binary stars.
The separate M-dwarf survey and its selection criteria are
discussed by \citet{EnCo03}.
Of all of these stars, 20 show large-amplitude variations indicative of 
previously unknown binary star systems,
and 11 additional stars show rms variations greater than 20\,m\,s$^{-1}$,
but probably not large enough to be due to binary stellar companions.
These stars represent good candidates for short-period planetary companions.
Figure~\ref{het_rms} shows a histogram of the velocity rms about the mean
of our time-series measurements for the remaining
142 stars having an rms of 20\,m\,s$^{-1}$ or less,
after any statistically significant linear velocity trends have been removed.
The time span of the data for each star ranges from one week to greater than
three years.  The median rms of the 123 stars in Figure~\ref{het_rms}
with rms of 13\,m\,s$^{-1}$ or less is 4.0\,m\,s$^{-1}$.
If we examine just the 112 stars with rms of 8\,m\,s$^{-1}$ or less (the bulk
of the sample), then both the mean and the median rms of the remaining
``stable'' stars are just 3.6\,m\,s$^{-1}$.   The peak of the histogram
is centered in the 2-3\,m\,s$^{-1}$ bin.  We note that a significant
contributing factor to the observed velocity rms is intrinsic to the stars
themselves.  We have {\itshape not} attempted to correct these rms values
in any way for velocity variations induced by stellar photospheric and
chromospheric activity (``jitter'').  In general, the stars for which we
have obtained the lowest rms are old, inactive stars with spectra of high
signal-to-noise ratio. Thus, Figure~\ref{het_rms} clearly demonstrates
that we have achieved routine radial velocity precision of 3\,m\,s$^{-1}$
or better with the HET-HRS in routine ``production mode''
on a large sample of a wide variety of stars over a substantial time period.
Indeed, a significant fraction of our stars are showing radial velocity
precision of 2\,m\,s$^{-1}$ or better, which is comparable to the precision
achieved for high SNR observations with the ESO VLT/UVES by \citet{KuEnRo03}.
Our 2-3\,m\,s$^{-1}$ precision is significantly better than that obtained
with CORALIE \citep{NaMaPe01} or ELODIE \citep{BaQuMa96}, and is directly
competitive with the 3\,m\,s$^{-1}$ precision which is achieved on the Keck
HIRES \citep{VoMaBu00}, AAT \citep{JoBuTi02}, or the 2\,m\,s$^{-1}$
demonstrated by HARPS \citep{PeMaQu04}.

\section{The planetary companion to HD\,37605}
\subsection{Characteristics of the host star}
HD\,37605 (HIP 26664, BD+05 985, SAO 113015, G 99-22) is a V=8.69 K0V star.
The Hipparcos \citep{Hipparcos} parallax of 23.32\,mas corresponds
to a distance of 42.9\,parsec, and gives an absolute magnitude of
$\rm{M_V} = 5.51$.  
The stellar parameters: effective temperature, T$_{\rm eff}$,
surface gravity, log\,$g$, and microturbulence, $\xi$ listed in Table~\ref{star}
were derived using MOOG \citep{Sn73} with atmospheres based on the
1995 version of the ATLAS9 code \citep{CaGrKu97} following the
procedure outlined in \citet{PaSnCo03} and briefly described below.
Within IRAF, we fit the 20 FeI and 12 FeII lines listed in Table 1 of
\citet{PaSnCo03} with Gaussian profiles to obtain equivalent widths.
The T$_{\rm eff}$ was derived by requiring that individual line abundances
be independent of excitation potential. Likewise,
$\xi$ was derived by requiring that individual line abundances be
independent of line strength. The log\,$g$ was determined by
forcing ionization equilibrium between neutral and singly ionized Fe.
We adopted solar log\,$\epsilon$(Fe) of 7.52 from \citet{AnGr89},
which yields an [Fe/H] for HD\,37605 of 0.39$\pm$0.06.
\begin{deluxetable}{lr@{$\pm$}l}
\tablecaption{Stellar properties for HD 37605 \label{star}}
\tablewidth{0pt}
\tablehead{
\colhead{Parameter} & \multicolumn{2}{c}{derived value}}
\startdata
T$_{\rm eff}$ & 5475 & 50K \\
log\,$g$ & 4.55 & 0.1 \\
$\xi$  & 0.8 & 0.2 km s$^{-1}$ \\
{\rm [Fe/H]} & 0.39 & 0.07 \\
\enddata
\end{deluxetable}

\subsection{HET Radial Velocity Observations}
\begin{figure}
\plotone{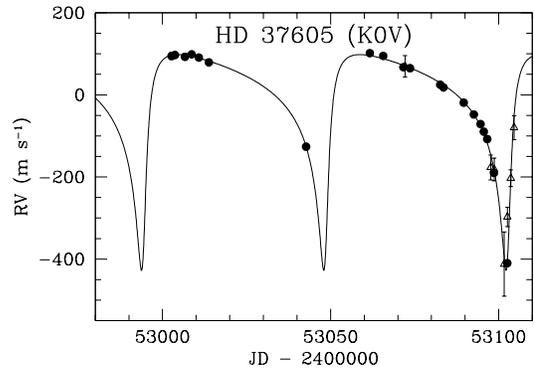}
\caption{Radial velocity data for HD~37605 from the HET HRS (filled circles)
and from the McDonald Observatory 2.1m Sandiford Cassegrain echelle
(open triangles).   The size of most of the HET error bars is smaller
than the size of the data points.  The best-fit combined Keplerian orbital
solution is plotted as solid line.   The rms residual of the HET/HRS data
from the orbital solution is 4.7\,m\,s$^{-1}$.}
\label{rvcurve}
\end{figure}

Our typical HET observing technique is to obtain about 4-5 initial
radial velocity
measurements over the course of 1-2 weeks, in order to search for short-period
``hot-Jupiter'' RV variability.  If a star appears stable on short timescales,
it is then scheduled for less frequent observations in order to search for
longer period orbits. For HD\,37605 ($V=8.7$), exposure times were 900 seconds,
giving a typical signal-to-noise ratio of about 250 per resolution element.
The initial set of observations of HD\,37605
were constant to within the observational error, but the next observation taken
about one month later showed a decrease of about 200\,m\,s$^{-1}$.
We then took advantage of the queue scheduled operations of the HET to
put the star back into the queue at high priority for frequent observations.
Following our request for frequent coverage, the HET Resident Astronomers
then obtained spectra every 3-4 days during the decrease in radial velocity.
Data from each night's observations were reduced and velocities computed the
following morning.   We were easily able to fit a new orbital solution as the
data from each night became available.   
It quickly became obvious that the radial velocity
minimum and periastron passage would occur just as the star was being lost
from the HET observability window; the last HET data points would be
obtained during evening twilight.  It was critical to attempt to obtain
nightly HET velocity measurements during this orbital phase, as the depth of
the velocity minimum would constrain the orbital eccentricity and the
K~velocity (and hence the companion minimum mass).   Since these were going
to be difficult HET observations, and it was quite possible that the HET would
no longer be able to observe the star before the periastron passage was
complete, we arranged for the McDonald Observatory 2.1m telescope observer
to obtain radial velocity measurements with the Sandiford Cassegrain Echelle
spectrograph \citep{MCSaBo93} and its I$_2$ cell during this interval. 
This instrument records the 5000 to 6000{\AA} region at resolving power of
60,000.  These 2.1m data would be of significantly lower velocity
precision due to the much lower signal-to-noise ratio and the inherent
mechanical and thermal instability of a large heavy Cassegrain spectrograph,
but they would certainly be adequate to cover this crucial interval when
the HET might not be able to observe the star.
The HET data for HD\,37605 are shown as filled dots in
Figure~\ref{rvcurve}, and the 2.1m Sandiford echelle data are shown
as open triangles.  The observed velocities from both telescopes are given
in Table~\ref{RVTable}. The velocities in this table have been corrected
for the different (and arbitrary) velocity zero points of the two instruments
used.  The simultaneous orbital solution to both data sets is also shown
in Figure~\ref{rvcurve}.  The rms residual of the HET/HRS data from
this orbital solution is 4.7\,m\,s$^{-1}$.   The rms residual of the 2.1m
Sandiford echelle data from the solution is 44.2\,m\,s$^{-1}$.
\begin{deluxetable}{rrrc}
\tabletypesize{\small}
\tablecaption{
Differential radial velocities for HD~37605 
\label{RVTable}
}
\tablewidth{0pt}
\tablehead{
\colhead{JD-2400000} & \colhead{velocity} &
\colhead{uncertainty} & \colhead{instrument} \\
\colhead {} & \colhead{(m\,s$^{-1}$)} & \colhead{(m\,s$^{-1}$)} & \colhead{}
}
\startdata
53002.6715 &    94.7 & 8.4 & HET HRS \\
53003.6853 &    97.1 & 7.8 & HET HRS \\
53006.6620 &    92.8 & 7.0 & HET HRS \\
53008.6641 &    98.6 & 6.2 & HET HRS \\
53010.8048 &    90.8 & 6.2 & HET HRS \\
53013.7940 &    79.3 & 6.6 & HET HRS \\
53042.7280 &  -126.4 & 7.0 & HET HRS \\
53061.6676 &   101.7 & 4.2 & HET HRS \\
53065.6468 &    94.8 & 4.8 & HET HRS \\
53071.6438 &    67.5 & 4.6 & HET HRS \\
53073.6382 &    64.9 & 4.2 & HET HRS \\
53082.6237 &    24.6 & 4.1 & HET HRS \\
53083.5954 &    18.2 & 5.6 & HET HRS \\
53089.5958 &   -19.0 & 4.0 & HET HRS \\
53092.5980 &   -47.9 & 4.1 & HET HRS \\
53094.5866 &   -71.2 & 3.3 & HET HRS \\
53095.5864 &   -89.6 & 3.5 & HET HRS \\
53096.5874 &  -107.7 & 6.7 & HET HRS \\
53098.5763 &  -190.0 & 4.2 & HET HRS \\
53102.5727 &  -409.9 & 5.5 & HET HRS \\ \hline
53072.1363 &      69.5 & 25.99 & 2.1m \\
53097.6345 &    -177.0 & 30.24 & 2.1m \\
53098.6555 &    -181.7 & 27.91 & 2.1m \\
53101.6647 &    -412.1 & 78.12 & 2.1m \\
53102.6034 &    -297.5 & 23.58 & 2.1m \\
53103.5928 &    -203.0 & 20.37 & 2.1m \\
53104.5972 &     -79.7 & 29.20 & 2.1m \\
\enddata
\end{deluxetable}

\subsection{Orbital Solution}
We performed a simultaneous fit of the HET-HRS and 2.1m Sandiford echelle
velocity data using the $Gaussfit$ \citep{MAJeMC94} software, which
uses a robust estimation method to find the combined orbital solution,
where the arbitrary velocity zero points of both
data sets are included as fit parameters.
The combined orbital solution is given in Table~\ref{Orbit},
and is shown as the solid line in Figure~\ref{rvcurve}.
The planet is in a highly eccentric orbit.   Its eccentricity of 0.737
is exceeded only by HD\,80606b (e = 0.93) and HD\,222582b (e=0.76).
The periastron distance of 0.070 {\scshape au} is large
enough for the planet easily to have avoided tidal circularization over
the lifetime of the system.
\begin{deluxetable}{llr@{$\pm$}l}
\tablecaption{Orbital parameters for the companion to HD~37605. \label{Orbit}}
\tablewidth{0pt}
\tablehead{\multicolumn{2}{c}{Parameter}
& \multicolumn{2}{c}{derived value}}
\startdata
P                    & [days] &             54.23      & 0.23 \\
T                    & [JD]   &              2452994.27 & 0.45 \\
K &       [${\rm m\,s}^{-1}$] &             262.9      & 5.5  \\
$e$     &                     &             0.737      & 0.010\\
$\omega$          & [degrees] &             211.6      & 1.7  \\ \hline
$f(m)$       & [solar masses] & $3.13\times10^{-8}$    &$5.9\times10^{-10}$\\
$m\sin i$&$[{\rm M}_{\rm Jup}]$&             2.84       & 0.26 \\
$a$         & [{\scshape au}] &             0.26       & 0.01 \\
\enddata
\end{deluxetable}

The shape of the observed radial velocity curve is very difficult to
explain by any mechanism other than orbital motion.  Nevertheless, we
have examined the available data to ensure that our interpretation of
the observed velocity variations as resulting from orbital motion is
indeed correct.   The Hipparcos photometry of HD\,37605 shows an
rms variation of only 0.0020 magnitudes, which is comparable to the
Hipparcos photometric precision.   Thus, the observed RV variations can
not be linked to any photometric variability of the primary star.
We have also searched for evidence of chromospheric variability of HD\,37605.
Normally, we would prefer to measure the Ca~II S index from the Ca~II H and K
lines, but the HET primary mirror coatings and the HET optical fibers
result in a short-wavelength cutoff of the HET HRS at around 4200{\AA}.
We do have one spectrum of HD\,37605 obtained with the McDonald Observatory
2.7m telescope 2dcoud\'e spectrometer, which includes the Ca~II H\,\&\,K lines.
From this spectrum we measure a McDonald S index \citep{PaSaCo02} of
$0.158\pm0.022$.  This ${\rm S}_{\rm McD}$ is comparable to the value found in
the old, inactive star $\tau$~Ceti (${\rm S}_{\rm McD}=0.166 \pm 0.008$), thus
indicating a low level of chromospheric activity for HD\,37605.
We have also examined the profiles of the H\,$\alpha$ absorption line
for any possible variability that might be related to stellar activity.
A careful alignment and normalization of the H\,$\alpha$ line profile from all
of our regular program spectra of HD\,37605 shows no variations beyond what is
expected from photon noise.
Thus we conclude that the observed radial velocity variations are indeed due
to a planetary-mass companion moving in a highly eccentric orbit.

\section{Conclusions}

We report the first discovery of an extra-solar planet using the Hobby-Eberly
Telescope and its High Resolution Spectrograph.
The planetary-mass companion to the metal-rich star HD\,37605 is in a
highly eccentric orbit.
We have demonstrated that the HET High Resolution Spectrograph can produce a
long-term velocity precision of 3\,m\,s$^{-1}$ or better over a multi-year
timespan.
The queue-scheduled operation of the Hobby-Eberly Telescope made the detection
and characterization of the orbit far simpler than it would have been with a
conventionally scheduled telescope.   Once the possible orbital motion was
detected, we were able to obtain immediate follow-up observations at the
particular orbital phases that were crucial for the orbit determination.
This highly flexible mode of telescope operation,
coupled with the large telescope aperture and the
extremely stable High Resolution Spectrograph, makes the HET-HRS an ideal
and extremely efficient and flexible 
facility for large surveys for extrasolar planetary systems.

\acknowledgements
The Hobby-Eberly Telescope (HET) is a joint project of the University of
Texas at Austin, the Pennsylvania State University,  Stanford University,
Ludwig Maximillians Universit\"{a}t M\"{u}nchen, and
Georg August Universit\"{a}t
G\"{o}ttingen.  The HET is named in honor of its principal benefactors,
William P. Hobby and Robert E. Eberly.
This material is based upon work supported by the National Aeronautics and
Space Administration under Grant NAG5-13206.
The authors are grateful the the University of Texas Hobby-Eberly Telescope
Time Allocation Committee for their generous allotment of observing time for
this program.

\end{document}